\documentstyle[11pt]{article}

\input epsf

\begin{document}

\titlepage

\def\Re{\mbox{Re}}
\def\a{\alpha}
\def\b{\beta}
\def\g{\gamma}
\def\d{\delta}
\def\e{\epsilon}
\def\x{\times}
\def\sss{s}
\def\om{\Omega}
\def\zee{z}
\def\k{\kappa}
\def\D{ {\cal D}}
\def\beq{\begin{equation}}
\def\eeq{\end{equation}}
\def\beqa{\begin{eqnarray}}
\def\eeqa{\end{eqnarray}}
\def\mxth{\mathsurround=0pt }
\def\xversim#1#2{\lower2.pt\vbox{\baselineskip0pt \lineskip-.5pt
\def\ialign{$\mxth#1\hfil##\hfil$\crcr#2\crcr\sim\crcr}}}
\def\simgr{\mathrel{\mathpalette\xversim >}}
\def\simle{\mathrel{\mathpalette\xversim <}}
\def\beqa{\begin{eqnarray}}
\def\eeqa{\end{eqnarray}}
\def\O{{\cal O}}
\def\L{{\cal L}}
\def\expt#1{\mathord{<}#1\mathord{>}}

\def\bi{\begin{itemize}}
\def\ei{\end{itemize}}
\def\i{\item}
\def\ra{\rightarrow}
\def\lag{Lagrangian}
\renewcommand{\thesection}{\Roman{section}. }
\renewcommand{\theequation}{\arabic{equation}}

\begin{flushright}
TUM-HEP-202/94\\
hep-th/9410205
\end{flushright}
\vspace{1ex}

\begin{center} \bf

S-DUAL GAUGINO CONDENSATION AND SUPERSYMMETRY BREAKING
      \rm
\vspace{1ex}

{\bf Z. Lalak$^\dagger$, A. Niemeyer$^\dagger$ and
H. P. Nilles$^{\dagger,\ddagger}$}

\vspace{0.5cm}
 $^\dagger$ {Physik Department} \\
       {\em Technische Universit\"at M\"unchen} \\
       {\em D--85748 Garching, Germany}

\vspace{0.5cm}
 $^\ddagger$ Max Planck Institut f\"ur Physik \\
  {\em Heisenberg Institut}\\
  {\em D--80805 Munich, Germany}

\vspace{1cm}

ABSTRACT
\end{center}
The principle of S-duality is used to incorporate gaugino condensates into
effective supergravity (superstring) Lagrangians. We discuss two
implementations of S-duality which differ in the way the coupling constant is
transformed. Both solve the problem of the runaway dilaton and lead to
satisfactory
supersymmetry breaking in models with a {\em single} gaugino condensate. The
breakdown of supersymmetry is intimately related to a nontrivial transformation
of the condensate under T-duality.

\vspace{1cm}

\begin{flushleft}
TUM-HEP-202/94 \\
September 1994
\end{flushleft}

\newpage

\subsection*{Introduction} 

Until now no satisfactory solution for realistic SUSY breakdown in superstring
inspired supergravity
models has been found. The main flaw of the models considered in the literature
is the difficulty of fixing the dilaton, the field which sets the value of the
gauge coupling at the unification scale, at a physically acceptable value.
Solutions proposed so far in the context of gaugino condensation involve
several gaugino condensates and an unnatural adjustment of the hidden matter
sector \cite{ccm}. However it seems possible to rectify the problem of the
runaway
dilaton in a much more fundamental way by invoking a new symmetry of the
constituent Lagrangian, the so-called S-duality.

Gaugino condensation in itself is inherently a field theoretical phenomenon and
it might be that S-duality gives the proper way of promoting it into
the string theoretical framework.
S-duality invariant effective purely dilatonic superpotentials have been
conjectured in \cite{fi} (however with no reference to gaugino condensation)
and recently reexamined by the authors of \cite{hm}. The general form of the
superpotential proposed in \cite{fi}, which was constructed specifically to fix
the vev of the dilaton, does not give a free theory in the weak
coupling limit. The authors of Ref.\ \cite{hm} note
that one can easily modify any effective
superpotential in such a way, that it vanishes asymptotically as $\Re S
\rightarrow \infty$ in any direction. Their one-condensate model
shows a realistic minimum for the dilaton but, unfortunately, SUSY is unbroken
at this minimum.

As gaugino condensation seems to be the most likely source of the
nonperturbative superpotential for the dilaton \cite{hpn}, it is important to
understand its role within S-dual string effective actions. Therefore we
study the dynamics of
supersymmetry breaking in a system containing both condensate and dilaton
degrees of freedom as well as the generic modulus $T$ (the breathing mode
common to all compactifications).

To construct the S-duality invariant form of the superpotential we note that
it has to contain a term originating from the gauge kinetic
term $(f {\cal W}_\alpha {\cal W}^\alpha)_F$ ($f$ is the gauge kinetic
function ($\Re f=1/g^2$) and the chiral superfield ${\cal W}_\alpha {\cal
W}^\alpha$ contains the gaugino bilinear as its lowest component) in the
constituent Lagrangian.
In order for this term to be invariant under S-duality one has to assume a
specific transformation law for the condensate superfield.

In its simplest realization, S-duality is an $SL(2,Z)$ symmetry generated by
$S \rightarrow 1/S, \; S\rightarrow S+i$. We shall discuss two physical
realizations of S-duality which differ in the way the coupling constant
transforms under the action of the first generator:

\begin{description}
\item[Type-I S-duality:] this is described by $S\rightarrow 1/S$ and $f
\rightarrow f$\\
 (or equivalently $g_{np}^2 \rightarrow g_{np}^2) $\footnote{$g_{np}$ denotes
 the  nonperturbative coupling constant (tree-level plus nonperturbative
 contributions).}.
\item[Type-II S-duality:] defined by the condition  $f\rightarrow 1/f$ (or
equivalently $g_{np}^2 \rightarrow 1/g_{np}^2$).
\end{description}

This second possibility demands the introduction of an additional field beyond
the gaugino condensate which could be called `magnetic condensate' $H$. Under
the type-II transformation $g_{np}^2 \rightarrow 1/g_{np}^2$ $H$ and $Y$ would
be exchanged. This is an interesting possibility, however, it is not obvious
whether an $N=1$ supersymmetric low-energy field theoretical description is
reliable in this case.

\subsection*{Type-I S-duality}

The simplest choice is to consider a tree-level duality invariant condensate.
This forces one to assume that the full nonperturbative gauge kinetic function
$f$ is also invariant under S-duality.

The correspondence to the weak coupling scattering amplitude calculations
demands that $f \rightarrow S$ as $S \rightarrow \infty$. A form of $f$ which
fulfills this requirement is

\beq
f=\frac 1{2\pi} \ln  (j(S)-744),\\
\eeq
where $j$ is the generator of modular functions of weight 0.
This means that the effective nonperturbative coupling constant is different
from $1/\sqrt {\Re S}$ which has some consequences we will discuss later.
Fig 1 shows the behavior of the nonperturbative coupling constant $g_{np}^2$
as a function of $S$.

\epsfbox[-80 0 500 210]{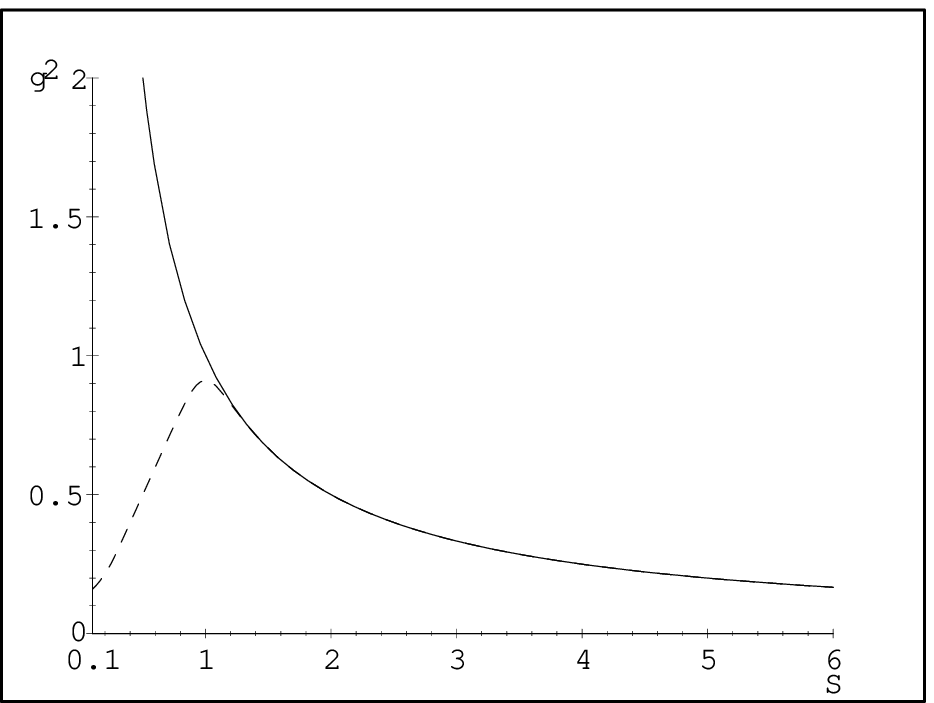}

{\small \em Fig. 1 - Nonperturbative coupling constant $g^2_{np}$ as the
function of $S$ in type-I models (dashed) vs $g^2$ given by $f=S$}

\vspace{0.3cm}

S-modular invariance demands that the superpotential transforms as a modular
form of weight $-1$ (assuming the standard K\"ahler term $-\ln (S+\bar S)$).
The simplest generalization of the Veneziano-Yankielowicz superpotential
\cite{vy} is then

\beq
W=\frac{Y^3}{\eta^2 (S)} (\frac 1{2\pi} \ln (j(S)-744) +
 3 b \ln (Y \eta^2 (T)/\mu) +c),\\
\eeq
with $Y^3={\cal W}_\alpha {\cal W}^\alpha$. For simplicity we assume $Y$ to
transform under
$T$-duality like a generic matter field of modular weight $-1$. The K\"ahler
function is then of the simple form

\beq
K=-\ln(S+\bar S) - 3 \ln (T+\bar T- Y\bar Y),\\
\eeq
where $b$ is the usual group theoretical constant, which we take to be $0.1$,
and $\mu$ is the order of magnitude where we expect the condensate $Y$ to form.
We take it to be $\mu=10^{-5}$ in Planckian units. We will denote the actual
vev of the condensate with $\mu^\prime$.
We adjust the constant $c$ in the way that for $S,T=1$ the gaugino
condensate assumes an expectation value of exactly $Y=\mu$. Note that $c$ can
be reabsorbed into the scale $\mu$.

This model gives rise to a scalar potential which becomes infinitely large as
$S \rightarrow \infty$ in a generic direction (and $S\rightarrow 0$, because of
S-duality). However, it is straightforward to verify that $V\rightarrow 0$ if
the fields go to infinity along the direction given by the condition $W_Y=0$.
Fig. 2 shows the scalar potential along this quasi-flat direction (because
$|W_Y|^2$ is the leading term in the scalar potential for small $Y$) for fixed
value of $T$. Running $S$ to infinity along this valley causes $Y$ to
vanish asymptotically. If we write the effective superpotential on the
hyper-surface $W_Y=0$, which means integrating out the condensate in the
usual way \cite{lt}, we get

\beq
W=\frac{const}{(j(S)-744)^{1/2 \pi b} \eta^2(S) \eta^6(T)}.
\eeq

\epsfbox[-80 0 500 210]{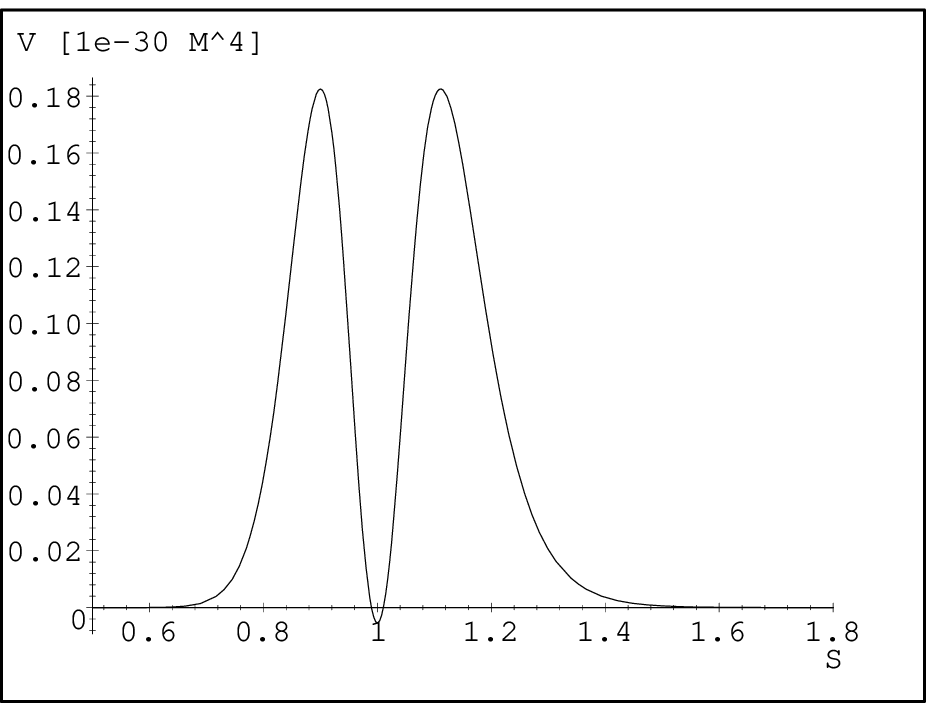}

{\small \em Fig. 2 - Scalar potential along the quasi-flat direction in type-I
models for a fixed value of $T$. The potential falls off in the weak coupling
regime}

\vspace{0.3cm}

This resembles the form conjectured in Ref.\ \cite{hm} except for the
different functional behavior in $j$ and most importantly the $T$-dependent
term (in \cite{hm} no modulus
field is included) in the denominator. It differs from the
type of superpotential proposed in Ref.\ \cite{fi}, where a regular
superpotential is required, whereas (in the context of gaugino condensation)
we see ourselves forced to admit singularities at
finite values of $S$ (though not on the real axis).

It is interesting to note that the weak-coupling limit of our model is
different from that of Ref.\ \cite{fi} and the one of Ref.\ \cite{hm}. In Ref.\
\cite{fi} (where gaugino condensates are not considered) $V\rightarrow
\infty$ for $S\rightarrow \infty$ regardless of the direction one follows in
field space. The authors of Ref.\ \cite{hm} on the other hand argue that for
gaugino condensates one should have
vanishing scalar potential in the weak-coupling limit and consider potentials
where $V\rightarrow 0$ for $S\rightarrow \infty$ because of inclusion of a
factor of $1/j$. This prescription gives uniformly vanishing scalar potential
in this limit regardless of the actual field content of the model. Our solution
gives $V\rightarrow \infty$ for a generic direction
but $V\rightarrow 0$ along the quasi-flat direction $W_Y=0$, which
corresponds to integrating out the condensate. We note at this point, that
there is no physical principle which enforces such a strong uniform suppression
of the potential as implied by the overall inverse power of $j$. Thus it might
be that the
requirement implied in Ref.\ \cite{hm} is too strong. The physical condition
which has to be imposed on the model is that the expectation value of the
gaugino condensate vanishes in the weak-coupling limit, as it happens in our
type-I model (as will be seen to happen also in our type-II model).

But because we cannot avoid singularities one could also argue that
the $1/\eta^2(S)$ prefactor needed to give $W$ the correct modular weight,
could be modified to be of the form $1/(\eta^2(S) P(j(S)^{1/3}))$ with $P$
being a polynomial. Including such a factor, e.g.

\beq
W=\frac{Y^3}{\eta^2 (S) j^{1/3}(S)} (\frac 1{2\pi} \ln (j(S)-744) +
 3 b \ln (Y \eta^2 (T)/\mu) +c).\label{mod2}\\
\eeq
makes the potential vanish in the weak coupling limit regardless of the
direction one follows in field space.

Minimization of the scalar potentials belonging to these two
models reveals that in both cases supersymmetry is broken (in contrast to
\cite{hm}, where the modulus $T$ is not included)\footnote{In Ref.\ \cite{hm}
supersymmetry breaking is achieved using multiple gaugino condensates.}.
 The minimum, which is
a global minimum, lies at $S=1, T\simeq 1.23, Y=\mu^\prime\simeq \mu$. The
supersymmetry breaking scale is determined by the auxiliary field of the
modulus $T$: $\expt{F_T}\simeq {\mu^\prime}^3/M$, whereas $\expt{F_S}=0$ and
$\expt{F_Y}\simeq {\mu^\prime}^4/M^2$. The cosmological constant is negative
and of the
order $V_0 \simeq -{\mu^\prime}^6/M^2$. The value of the nonperturbative gauge
coupling constant at this minimum is $g^2=0.91$.

\subsection*{Type-II S-duality}

The type-II implementation of S-duality takes into account the fact that the
gaugino sector of the theory might not close under the S-duality
transformation. To write down a model which is invariant one has to include an
additional sector, the `magnetic condensate', which is supposed to represent
the dual phase of the theory \cite{vw}.

The simplest toy model which illustrates the idea of type-II S-duality is given
by

\beq
K = -\ln (S+\bar S) - 3 \ln (T+\bar T-Y \bar Y - H \bar H),
\eeq

\beq
W= \frac 1{\eta^2(S)} (Y^3 S + H^3/S + 3 b Y^3 \ln \frac{Y \eta^2(T)}{\mu}
+ 3 b H^3 \ln \frac{H \eta^2(T)}{\mu}+Y^3 H^3/\mu^3).
\eeq

This model does not exhibit a full $SL(2,Z)$-symmetry, but only the
strong-weak-coupling duality

\beq
f=S \rightarrow 1/S, \qquad Y \leftrightarrow H.
\eeq

In principle one could with some effort promote this symmetry to a full
$SL(2,Z)$, but for illustration of our statements we choose this simple model,
especially because both real and imaginary parts of $S$ already become fixed
even with this smaller symmetry.

The scalar potential possesses a (although rather hard to find) minimum close
to
$S=1, T=0.560, H=Y=\mu^\prime= 3.64\,10^{-2}\, \mu $. It turns out that at the
minimum supersymmetry is broken, with the magnitude of SUSY breaking again
determined by $\expt{F_T} \simeq {\mu^\prime}^3/M$, where $\mu^\prime$ is
the dynamically determined value of the condensate at the minimum
(in the previous type-I examples we adjusted (using $c$) $\mu^\prime$ to be of
the phenomenologically reasonable value $10^{-5}$). The cosmological constant
is again negative and of the order $\simeq -{\mu^\prime}^6$.

Of course it would be interesting to obtain an effective superpotential
containing only dilaton and modulus as we did in our type-I model. One
possibility would be to integrate out $Y$ and $H$-fields using the standard
conditions $W_Y=W_H=0$. These conditions are the equations of motion for $Y$
and $H$ in global SUSY and also mean that supersymmetry is unbroken in the
naive global limit. Integrating out $Y$ and $H$ is valid because they are
much heavier than $S$ and $T$, which can be read off the Hessian at the
minimum and because their contribution to SUSY breaking is negligible. The
procedure of integrating out using the relations
$W_Y=W_H=0$ is valid only if $Y$ and $H$ stay small with respect to the Planck
scale, because only then the terms $|W_Y|^2$ and $|W_H|^2$ are the dominant
contributions to the scalar potential. These are given by

\beqa
W_Y& =&\frac {3 Y^2}{\eta^2(S)} (S+3b\ln \frac {Y\eta^2(T)}\mu + b + \frac{H^3
\eta^6(T)}{\mu^3})\\
W_H& =&\frac {3 H^2}{\eta^2(S)} (\frac 1S+3b\ln \frac {H\eta^2(T)}\mu + b +
\frac{Y^3 \eta^6(T)}{\mu^3})\label{eq10}
\eeqa

One cannot solve both conditions simultaneously in an analytical way.
Since we are interested in the nature of the weak-coupling regime in this
model, we solve the above equations asymptotically for large $S$.
It is easy to see that in order to preserve $W_Y=0$, $Y$ has to go
exponentially to 0 as
$S\rightarrow
\infty$. Inserting this result into the second equation one obtains
two solutions: $H\rightarrow \exp(-\alpha S)$ with a sufficiently large
$\alpha$
(say 1), or $H$ staying finite and always exactly cancelling the bracket in
(\ref{eq10}), going ultimately to $\exp(-1/3) \mu/\eta^2(T)$. In order to use
these
relations we had to assume $Y, H$ to be small, which we consistently get
in both solutions for large $S$. After expressing the superpotential in
terms of $W_Y, W_H$

\beq
W=Y W_Y+H W_H- \frac 1{\eta^2(S)} (3b(Y+H)+Y^3 H^3 \eta^6(T)/ \mu^3)\\
\eeq
one can see that for $S\rightarrow \infty$ the superpotential in the first case
goes to zero exponentially, but blows up in the second case ($1/\eta^2(S)$
behaves asymptotically like $\exp(\pi S/12)$). This implies in
turn that the scalar potential $V$ for large $S$ becomes 0 in the first or
infinitely large in the second case, in contrast to
our type-I model. Taking the usual conditions for eliminating condensates one
can therefore arrive at different effective superpotentials.

However, this could be a very model dependent property. If
one wishes to have vanishing scalar potential in the weak
coupling limit, one could always include e.g. an $1/j(S)$ factor in the
superpotential. In addition we would like to point out that we only considered
the direction along which SUSY
is not broken by the auxiliary fields of the condensates.

\subsection*{Discussion and Conclusions}

Our analysis shows that it is possible to have realistic SUSY breaking in
type-I models with
a single gaugino condensate and without matter fields by invoking
S-duality. However, S-duality does not seem to solve the problem of the
cosmological constant, which is negative in both types of models. It
can be shown that even small non-S-dual perturbations of the superpotential
preserve the nice features of the above models \cite{zap}. Maybe they could be
used to cancel the cosmological constant.
\\Considering
more general modular invariant functions instead of $j^{-1/3}$ in (\ref{mod2})
might lead to a change of the magnitude of the supersymmetry
breaking scale. If the modulus $T$ is absent from the theory, the minima are
still well defined, but SUSY stays unbroken.

Note that the analysis for type-I models is based on the assumption that the
condensate
does not transform under S-duality. In principle it would be possible to assume
a nontrivial transformation behavior for the $Y$-field. However, one is forced
to constrain oneself to $Y$ and $f$ transforming as modular forms of opposite
weight, because with the minimal choice of variables in the superpotential ($S,
Y$) it is impossible to cancel an arbitrary transformation law in the
term $f Y^3$. In order to have $T$ transforming trivially under S-duality
(as it should at tree-level), one has to include appropriate
$\eta$-factors in the logarithm in the K\"ahler function, if $Y$ does transform
as some modular function of $S$. However, one can then see that it is easy to
redefine
$Y$ and $f$ by shifting some power of $\eta$-functions from one to the other,
so that one gets back to the form we proposed above.

\vspace{0.5cm}

For our type-II model several desirable properties are the same as for
the type-I models:
SUSY is again broken at a reasonable minimum, one observes no
supersymmetry breaking if one does not include the modulus $T$ and the scale of
SUSY breakdown is similar in both types of models as is the cosmological
constant. The weak-coupling limit however, is
different. Type-I models generically lead to vanishing scalar potential in the
limit $g^2 \rightarrow 0$ after eliminating the condensate through its
equations of motion. This is not the case for our type-II model, as the
integration conditions have different solutions with different asymptotic
behavior, but there seems to be no argument why the potential has to vanish in
the $g^2\rightarrow 0$ limit,
because asymptotically the gaugino condensate {\em does} vanish, as required.
The magnetic condensate $H$ may not vanish together with the gaugino
condensate and therefore there is no a priori quasi-flat direction in this
model in the weak-coupling regime. This also shows that the usual conditions to
eliminate condensates might have to be revised.

\vspace{0.5cm}

We have not considered stringy one-loop corrections here. These usually
introduce a mixing of $S$ and $T$ in the K\"ahler function, so that $T$ cannot
transform trivially under S-duality any longer. One could imagine that this
problem can be solved by introducing two new orthogonal fields $S^\prime,
T^\prime$ which do not mix under their respective duality transformations. This
would be in analogy to the usual redefinition of the dilaton in $T$-modular
invariant one-loop-gaugino condensation models.

\vspace{0.5cm}
We have demonstrated that the idea of S-duality applied to gaugino condensation
induced stringy superpotentials naturally leads to potentials with well-defined
supersymmetry breaking minima. We have proposed two different ways of
incorporating S-duality into the low-energy effective Lagrangian, which differ
in the way the coupling constant is transformed, but both lead to physically
satisfying solutions while employing only a single gaugino condensate. We feel
that this solution to the long-standing problem
of runaway dilaton vacua is more attractive than the traditional methods
relying on multiple condensates. In addition we have shed some light on the
necessity of having a vanishing scalar potential in the weak-coupling limit.

\subsection*{Acknowledgments}
This work was supported by the Deutsche Forschungsgemeinschaft and EC grants
SC1--CT92--0789 and SC1--CT91--0729. Z.L. has been
supported by the A. von Humboldt Fellowship. A. N. has been
supported by a PhD scholarship from the Technical University of Munich. We
would like to thank Andre Lukas and Peter Mayr for useful discussions.

\hspace*{.5cm}

\end{document}